\documentclass[a4paper]{mem}
\usepackage{natbib}
\usepackage{graphicx}
\usepackage[a4paper]{hyperref}
\idline{1}{1}
\begin{document}

   \title{Short-lived isotopes and $^{23}$Na production in low mass AGB Stars}

   \author{S. Cristallo \inst{1}, 
           R. Gallino \inst{2},
           O. Straniero \inst{1},
           L. Piersanti \inst{1},
           I. Dom\'inguez \inst{3}}


   \institute{Osservatorio Astronomico di Collurania, 
              via Mentore Maggini, 64100 Teramo, Italy; \email{cristallo@oa-teramo.inaf.it} 
              \and  Dipartimento di Fisica Generale, Universit\'a di Torino e Sezione INFN di Torino,
             via P. Giuria 1, 10125 Torino, Italy
             \and Departimento Fisica Teorica y del Cosmos, Universidad de Granada, 18071 Granada, Spain}

   \abstract{We discuss the synthesis of  
   some short-lived isotopes and of $^{23}$Na in thermally pulsing AGB
   stars with initial mass of 2 M$_{\odot}$ and two different metallicities ($Z$=1.5$\times$10$^{-2}$, corresponding
   to the metal amount in the present Sun, and $Z$=10$^{-4}$),
  representative of disk and halo stars, respectively.\\
   The different nucleosynthesis channels are illustrated in some details.
   As previously found, the 
   $^{13}$C formed after each third dredge up episode is usually completely consumed by $\alpha$ captures before the onset of
    the subsequent thermal pulse, 
   releasing neutrons. This is the most efficient neutron source in low mass AGB stars, 
   and the resulting s-process 
   nucleosynthesis is at the origin of
    the solar main component. However,
   in the solar metallicity model, we find that the temperature of the first formed $^{13}$C pocket
    remains too low during the interpulse 
   and the $^{13}$C is not completely burnt, being partially engulfed in the convective zone
    generated by the
   following thermal pulse. Due to the rapid convective mixing in this zone,
    the $^{13}$C is exposed to a larger
    temperature and a nucleosynthesis characterized by a relatively high neutron density develops.
    The main effect is 
   the strong enhancement of isotopes located beyond some critical branching in the neutron-capture path,
    like $^{60}$Fe, otherwise
   only marginally produced during a {\it standard} s-process nucleosynthesis. 
      
    --
                short-lived isotopes
   }
   \authorrunning{S. Cristallo et al.}
   \titlerunning{$^{23}$Na and short-lived isotopes in low mass AGB Stars}
   \maketitle

\section{Introduction}

Low mass AGB Stars (1$<$$M$/M$_{\odot}$$<$3) are among the most important polluters of the Milky Way,
 because of the 
strong winds eroding their chemically enriched envelopes. 
In the AGB phase, the material processed during the alternating series of H and He burnings is recurrently mixed within the 
convective zones generated by thermal pulses (TPs) \cite{sh65} and then partially mixed with the surface by the so called third dredge
 up (TDU) episodes. 
As a matter of fact, MS, S, C(N) and some post-AGB stars are enriched in C and s-process elements
and the spectroscopic detection of unstable Tc demonstrates that the s process is 
at work in these stars \cite{merr52}. \\
   \begin{figure}
   \centering
   \resizebox{\hsize}{!}{\includegraphics{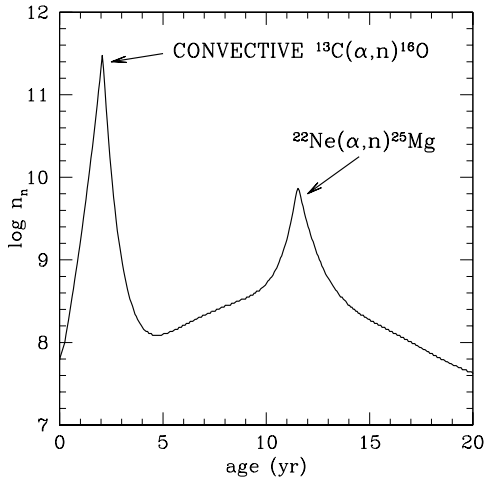}}
      \caption{Peak neutron density evolution after the ingestion of the $^{13}$C-pocket formed after the first TDU
      episode (solar metallicity model).} 
         \label{Fig1}
   \end{figure}
Besides $^{99}$Tc, a number of short-lived isotopes are produced during 
the s-process, among which $^{41}$Ca, $^{60}$Fe and $^{107}$Pd. Moreover,
$^{26}$Al is an important product of the H-shell burning, by proton capture on
$^{25}$Mg. In this work we want to study the impact of an improved version of
the FRANEC code on the production of short-lived radioactivities,
and discuss the challenging  hypothesis  that the protosolar nebula was
polluted by the winds of a close-by AGB star (see \cite{was06} for a detailed
review of this argument).
A major improvement in the stellar evolution code was the
introduction of a physical algorithm for the treatment of the
convective/radiative interfaces \cite{cris01} (see Section \ref{c13conv}). In such a way we succeeded to overcome
the discontinuity in the thermal gradients that otherwise
would result at the border of the H-rich convective envelope and the inner
radiative C and He-rich intershell region. Furthermore, the partial diffusion of
protons in the top layers of the He shell gives naturally rise to the
subsequent formation of a primary $^{13}$C pocket, partially overlapped with
an outer $^{14}$N pocket, followed by a further minor $^{23}$Na pocket, both of
primary origin.
The new models presented here are for AGB stars of initial mass $M $ = 2
M$_{\odot}$ and two metallicities (Z=1.5$\times$10$^{-2}$, Z=10$^{-4}$), representative of disk and halo stars, respectively. 
In Section \ref{c13conv} we discuss some key characteristics of the self-consistent
formation of the $^{13}$C pocket and of the $^{14}$N pocket along the AGB phase.
In Section \ref{radioact} we examine in detail the impact of the new models on the
production of short-lived radioactivities, while in Section \ref{sodium} we reanalyze 
the production of $^{23}$Na in AGB stars. 

\section{The cool $^{13}$C pocket}
\label{c13conv}

Two neutron sources have been identified as operating in thermally pulsing AGB stars. Concerning low mass AGBs,
 the main contribution to the s-process nucleosynthesis
comes from  the $^{13}$C($\alpha$,n)$^{16}$O 
reaction, activated during the interpulse phase \cite{stra95}, while a minor additional exposure is provided by 
the marginal activation of the
$^{22}$Ne($\alpha$,n)$^{25}$Mg reaction at the base of the convective zone generated by a TP. 
In our models, the introduction of an exponentially decaying
profile of the average velocity of the turbulent eddies implies
two important by-products. The first is a more efficient TDU, 
the total amount of  dredged up material being about the double of the one obtained without imposing a smoothed 
velocity profile.
The second by-product, very important for the nucleosynthesis, is  the diffusion of a small amount of protons 
($\sim$5$\times$10$^{-5}$ M$_{\odot}$)
in the underlying intershell region, mainly composed by $^4$He (80\%) and $^{12}$C (20\%). 
Later on, when the temperature in the contracting intershell region grows up,
these protons are  captured by the abundant $^{12}$C, leading to the formation of a tiny $^{13}$C pocket.
Then, if the temperature attains $T_8\sim$ 0.9 ($T_8$ is the temperature in units of 10$^8$ K), 
the $^{13}$C burns, releasing neutrons and activating the 
s-process nucleosynthesis. Such a condition is normally reached during the interpulse phase 
and the $^{13}$C is fully consumed before the onset of the subsequent TP \cite{stra95}.
The resulting nucleosynthesis is characterized by a rather low neutron density, 
never exceeding 10$^7$ neutrons/cm$^3$ for the solar metallicity model, or 10$^8$ neutrons/cm$^3$ for the Z=10$^{-4}$ model.
In our solar metallicity model, a typical neutron exposure
reached during a radiative $^{13}$C burning (weighted over the whole pocket) is $\delta\tau\sim$ 0.25 mbarn$^{-1}$.
The s-rich pocket is then engulfed by the convective zone generated by the TP and, later on, dredged up
by the next TDU.
The partial activation of the $^{22}$Ne($\alpha$,n)$^{25}$Mg reaction during the thermal pulse 
provides an additional minor neutron exposure.
In this case a neutron burst with high peak neutron density  
(10$^{10}$-10$^{11}$ neutrons/cm$^3$, depending on the maximum temperature reached at the base 
of the convective zone) is obtained.\\
%
As discussed in previous papers (see {\it e.g.} \cite{cris04}),  
in the solar metallicity model,
at the beginning of the AGB phase, the $^{13}$C pocket formed after the first TDU episode is not fully consumed   
and is partially engulfed by the convective shell generated 
by the next thermal pulse. 
The $^{13}$C burning takes place 
at the bottom of the convective shell
($T_8\sim$ 1.6) producing a maximum neutron density 
$n_n$=3.3$\times$10$^{11}$ neutrons/cm$^3$, about 30 times larger
than the neutron density released by the subsequent $^{22}$Ne neutron burst
(see Fig. \ref{Fig1}). These numbers refer to the maximum neutron density attained at the bottom of the
convective shell generated by the TP, while the average neutron density within the shell is one order 
of magnitude lower (see \cite{ga88}).
The contribution to the nucleosynthesis from this anomalous convective $^{13}$C burning reveals interesting peculiarities.
In particular, some branchings, which remain closed during a standard radiative $^{13}$C burning and are marginally
activated during the $^{22}$Ne($\alpha$,n)$^{25}$Mg burning, are now open.
In a very short temporal step ($\Delta T<3$ years, see Fig. \ref{Fig1}) we obtain a consistent production
of neutron-rich isotopes normally bypassed by the standard radiative $^{13}$C s-process, among which  
$^{60}$Fe (see Section \ref{radioact}), $^{86}$Kr, $^{87}$Rb and $^{96}$Zr.

\section{Radioactive Isotopes}
\label{radioact}

\begin{figure*}
   \centering
   \includegraphics[width=11cm]{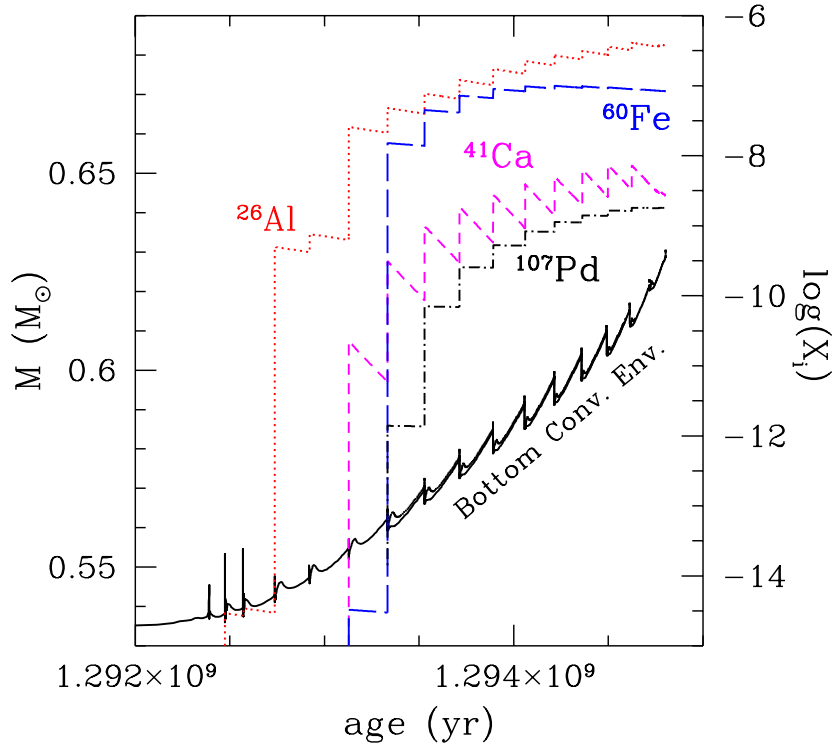}
      \caption{TP-AGB surface evolution of $^{26}$Al, $^{41}$Ca, $^{60}$Fe and $^{107}$Pd (solar metallicity model).}
         \label{Fig4}
   \end{figure*}

In this Section, we want to verify the hypothesis 
that a single low mass AGB star had contaminated the protosolar nebula right before its collapse (as suggested by \cite{was06}),
by comparing the measured early solar system abundances of different short-lived isotopes
to the predicted surface composition of our solar metallicity model.
Before explaining the procedure we follow, we describe the nucleosynthetic processes
responsible of stellar surface enrichments of $^{26}$Al, $^{41}$Ca, $^{60}$Fe and $^{107}$Pd.\\
Ground state $^{26}$Al has a half-life of 7.16$\times$10$^5$ years and is efficiently produced in the 
H-burning shell by proton captures on $^{25}$Mg. 
The surface is firstly enriched of freshly synthesized $^{26}$Al when the convective 
envelope partially erodes the H-burning shell (first bump in the $^{26}$Al profile in Fig. \ref{Fig4}) and, later on, as a consequence
of TDU episodes.
Note that a stellar layer exposed to a neutron flux is substantially depleted in $^{26}$Al, 
because of the large neutron-capture cross section of this isotope. 
This happens during both the radiative and convective $^{13}$C burnings as well as during the convective $^{22}$Ne burnings.
When the temperature at the bottom of the convective shells generated by TPs
exceeds $T_8\sim$2.7, the $^{26}$Al(n,p)$^{26}$Mg reaction efficiently destroys the $^{26}$Al.
In our solar metallicity model this condition is attained only toward the end of the AGB phase,
so that $^{26}$Al left in the H-burning shell ashes is largely preserved (with the exception of the
thin layer where the radiative $^{13}$C burning takes place). 
For this reason we found a substantial surface enhancement of $^{26}$Al.\\
The appearance at the surface of $^{41}$Ca and $^{60}$Fe is simultaneous 
with the occurrence of the first TDU episode (see Fig. \ref{Fig4}). 
As already recalled, the high $^{60}$Fe production we obtain 
is a direct consequence of the engulfment into the
convective zone generated by the next TP
of the $^{13}$C contained in the first
$^{13}$C-pocket.
The final surface isotopic ratio $^{60}$Fe/$^{56}$Fe we obtain in the solar metallicity model 
is equal to 6.5$\times$10$^{-5}$, about 
15 times higher then the value obtained in the case of the 
{\it standard} radiative $^{13}$C-burning (see for a comparison Table 4 of \cite{was06}).
On the contrary, in the $Z$=10$^{-4}$ model, no cool $^{13}$C pockets are found.\\
Concerning $^{41}$Ca, 
the equilibrium value ($^{41}$Ca/$^{40}$Ca$)\sim$10$^{-2}$ is normally 
attained during an efficient s-process nucleosynthesis. Thus, an excess of $^{41}$Ca
can only be obtained in the case of a very small neutron flux. This occurs when the
$^{13}$C left in the He-rich intershell by the H-burning during the Early-AGB phase is burnt
 by $\alpha$ captures at the beginning of the thermally-pulsing phase,
before the occurrence of the first TDU episode. \\
The $^{107}$Pd is mainly synthesized during the {\it standard} radiative $^{13}$C burning inside $^{13}$C-pockets, when isotopes are
exposed to large neutron exposures.  
The more the number of radiative $^{13}$C burning episodes is, the larger the $^{107}$Pd abundance in the envelope is.\\
Under the hypothesis that the protosolar nebula was
contaminated by a single AGB star, we follow the procedure described in \cite{was06}: 
we adjust the isotopic ratios at the end of our model by assuming a 
dilution factor ({\it f}) and a delay time ($\Delta$) between the end of the AGB phase and 
the beginning of the pollution process (see Table 2). We use the $\Delta$ assumed in \cite{was06} (see their Table 5) and we
tune the dilution factor in order to match the Pd isotopic ratio. 
We find too low Al, Ca and Fe isotopic ratios with respect ESS measurements. 
The situation is completely different if we analyse the {\it "relative"} isotopic enhancements just after the  2$^{nd}$ TDU: at that epoch
the internal layers has suffered a unique low neutron exposure due to the
convective $^{13}$C burning ($\delta\tau\sim$ 0.05 mbarn$^{-1}$), avoiding in such a way the later contribution 
from radiative $^{13}$C burnings (which strongly
increase the Pd isotopic ratio). 
We report in Table 2 (column 4) the isotopic ratios after the 2$^{nd}$ TDU, obtained by adjusting again the
two free parameters. 
We found an agreement for Ca, Fe and Pd isotopic ratios, but a too low $^{26}$Al/$^{27}$Al turns out.
We could however invoke the occurrence of Cool Bottom Process \cite{nol}, which
produces $^{26}$Al at the base of the convective envelope enhancing its surface abundance.\\
Our guess is that the astrophysical source polluting the ESS was a lower mass AGB star (1.3$\div$1.5 M$_{\odot}$):
in that case the number and the strength of the TDU episodes would decrease and the final isotopic distribution 
could be similar to the one obtained after the 2$^{nd}$ TDU episode of our $M$=2 M$_{\odot}$ (provided the occurrence of the first convective
$^{13}$C burning, needed to increase the $^{60}$Fe abundance).
This hypothesis has however to be verified by a full stellar model.

\section{$^{23}$Na}
\label{sodium}
\begin{figure}
   \centering
   \resizebox{\hsize}{!}{\includegraphics{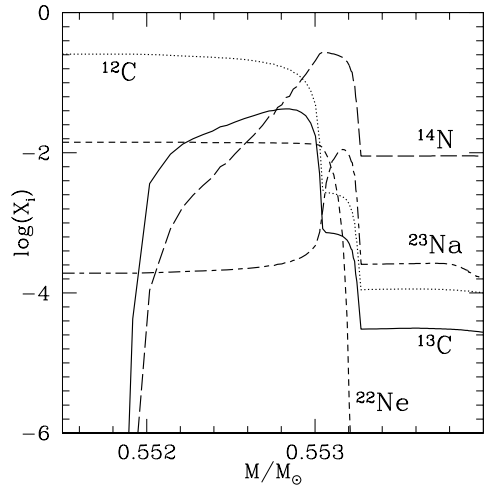}}
      \caption{Abundances profiles in the $^{13}$C pocket formed after the first TDU episode in the
solar metallicity model. 
}
         \label{Fig2}
   \end{figure}

In thermally pulsing AGB stars, sodium is synthesized by proton captures on $^{22}$Ne.
We recall that the He-rich intershell is progressively enriched in $^{14}$N, left in the ashes 
of the H burning, and then fully converted to $^{22}$Ne in the convective zone powered by the He-shell flash,
by the sequence $^{14}$N($\alpha$,$\gamma$)$^{18}$F($\beta^-$)$^{18}$O($\alpha$,$\gamma$)$^{22}$Ne.
\begin{table*}[htb]                                   
\begin{center}                                         
\begin{tabular}{|c|c|c|c|}
\hline 
Isot. Ratio & ESS Inventory & END AGB (11$^{th}$ TDU) & 2$^{nd}$ TDU\\
& & {\it f}=1.6$\times$10$^{-4}$ $\Delta$=0.68 Myr & {\it f}=1.9$\times$10$^{-2}$ $\Delta$=0.25 Myr\\
\hline 
$^{26}$Al/$^{27}$Al & 5$\times$10$^{-5}$ & 3.3$\times$10$^{-7}$ & 1.0$\times$10$^{-5}$\\
$^{41}$Ca/$^{40}$Ca & 1.5$\times$10$^{-8}$ & 5.2$\times$10$^{-11}$ & 1.5$\times$10$^{-8}$\\
$^{60}$Fe/$^{56}$Fe & (2$\div$20)$\times$10$^{-7}$ & 4.4$\times$10$^{-9}$ & 1.9$\times$10$^{-7}$\\
$^{107}$Pd/$^{108}$Pd & 2.0$\times$10$^{-5}$ &  2.0$\times$10$^{-5}$ & 2.0$\times$10$^{-5}$\\
\hline
\end{tabular} 
\end{center}
\caption{Measured and predicted surface isotopic ratios involving short-lived isotopes. (see text for details).} 
\end{table*}   
In Fig. \ref{Fig2} we plotted, for the solar metallicity model, the chemical profiles 
of some trace isotopes in the region around the $^{13}$C pocket for a typical interpulse of 
the solar-metallicity model. 
The $^{13}$C pocket (solid line) partially overlaps with a more external $^{14}$N pocket 
(long-dashed line). The maximum of the $^{14}$N coincides with the region where 
the protons diffused from the convective envelope  
at the epoch of the TDU were abundant enough to allow a full CN cycle.  
This is also the region where the $^{23}$Na (short-dashed-long-dashed line) has been efficiently produced 
by the $^{22}$Ne(p,$\gamma$)$^{23}$Na  reaction.  Indeed, at the end of the previous thermal pulse, 
the mass fraction of $^{22}$Ne in the top layer of the He-rich intershell is of the order of 1.5 10$^{-2}$ 
and, around the $^{14}$N maximum, this $^{22}$Ne is almost completely converted into $^{23}$Na,
leading to the formation of a tiny
$^{23}$Na pocket. 
In the solar metallicity model, the $^{23}$Na production is dominated by this mechanism. In the Z=10$^{-4}$ model, 
a second source of $^{23}$Na is represented by neutron capture on $^{22}$Ne, which occurs both
during the radiative $^{13}$C burning and  
during the convective $^{22}$Ne burning. For this model, these additional sources contribute to 13\% and 35\% of the total sodium 
production, respectively.\\ 
%
%
While our solar metallicity model shows a negligible $^{23}$Na surface enrichment (+25\%, equivalent to 0.1 dex) during the AGB phase, 
at the end of the Z=10$^{-4}$ model we found a sodium surface enhancement of more than a factor 10
([Na/Fe]$\sim$1). It is therefore worth to analyse the effects of
varying some key reaction rates affecting the $^{23}$Na production in the low metallicity model.
Recent measurements of $^{22}$Ne(p,$\gamma$)$^{23}$Na \cite{hale1} and $^{23}$Na(p,$\gamma$)$^{24}$Mg 
reaction rates \cite{hale2} significantly reduce the values quoted in the NACRE compilation \cite{Angulo}
around $T_8$=1. 
The reduction of the first reaction rate depresses the abundance peak
of the $^{23}$Na pocket; this effect  is marginally counterbalanced
by the reduction of the second reaction rate.  
We then recalculate a series of TPs in the Z=10$^{-4}$ model with rates from (\cite{hale1},\cite{hale2}) and we compare the resulting
intershell sodium mass fraction with the one obtained with the NACRE rates \cite{Angulo}: 
a variation of about $-$30\% is found. Moreover, we verified that the correction of a misprint in the 
$^{23}$Na(n,$\gamma$)$^{24}$Na reaction rate (5 keV data tabulated in \cite{bao}, see contribution from Heil et al. in these
Proceedings) implies a further decrease by about 10\%. This correction
makes  the contribution of the neutron capture on $^{22}$Ne during radiative $^{13}$C burning negligible, while
it does not affect the $\sim$23 keV neutron capture occurring in the convective $^{22}$Ne burning. 
As a whole, we estimate that with these changes of the relevant reaction rates the final overabundance of sodium will be [Na/Fe]$\sim$0.7.

\section{Conclusion}
In this paper we study how short-lived isotopes are created in a low mass ($M$=2 M$_{\odot}$) AGB star, 
showing the strong effects a single convective $^{13}$C burning episode implies. 
A possible explanation
for the ESS short-lived isotopes has been proposed, this however implying strong {\it ad hoc} assumptions.\\
In addition, we present detailed nucleosynthetic results concerning the production of $^{23}$Na in low mass AGB stars
at different metallicities (Z=1.5$\times$10$^{-2}$ and Z=10$^{-4}$), representative of disk and halo stars respectively.
Different processes responsible for the $^{23}$Na nucleosynthesis have been analyzed in detail, stressing the importance of 
varying key nuclear reaction rates.

\bibliographystyle{plain}

\begin{thebibliography}{9}

\bibitem {Angulo} 
Angulo, C. et al. 1999, Nucl. Phys. A, 656, 3 
\bibitem {bao}
Bao, Z.Y., \& K\"appeler, F. 2000, Atom. Data Nucl. Data Tables, 76, 70
\bibitem {cris01}
Cristallo, S. et al. 2001, Nucl. Pyis. A, 688, 217 
\bibitem {cris04}
Cristallo, S. et al. 2004, Mem. SAIt, 75, 676 
\bibitem {ga88} 
Gallino, R. et al. 1988, ApJ, 334, 45L
\bibitem {hale1} 
Hale, S.E. et al. 2002, Phys. Rev. C, 65, 015801
\bibitem {hale2} 
Hale, S.E. et al. 2004, Phys. Rev. C, 70, 045802
\bibitem {merr52}
Merrill, P.W.  1952, ApJ, 116, 21
\bibitem {nol}
Nollett, K.M. et al. 2003, ApJ, 582, 1036
\bibitem {sh65}
Schwarzschild, M. \& H\"arm, R. 1965, ApJ, 142, 855
\bibitem {stra95}
Straniero, O. et al. 1995, ApJ, 440L, 85S 
\bibitem {was06}
Wasserburg, J.J. et al. 2006, Nucl. Phys. A, {\it in press}

\end{thebibliography}

\end{document}